\documentclass{article}

\usepackage{arxiv}

\usepackage[utf8]{inputenc} 
\usepackage[T1]{fontenc}    
\usepackage{hyperref}       
\usepackage{url}            
\usepackage{booktabs}       
\usepackage{amsfonts}       
\usepackage{nicefrac}       
\usepackage{microtype}      
\usepackage{lipsum}
\usepackage{graphicx}
\usepackage{svg}
\usepackage{mathtools}
\usepackage{multirow}
\usepackage{amssymb}
\usepackage[export]{adjustbox}
\usepackage[numbers, sort&compress, comma, square]{natbib}

\title{Probabilistic Safety Analysis using Traffic Microscopic Simulation}

\author{
  Carlos Lima Azevedo\thanks{This work was carried out for Carlos PhD Thesis at the National Laboratory for Civil Engineering, Portugal, 2011-2014. This document is a revised and extended version of the paper presented by the authors at the 94th Annual Meeting of the Transportation Research Board, Washington DC, USA, 2015, with the same title.} \\
  Department of Management Engineering\\
  Technical University of Denmark\\
  Anker Engelunds Vej 1\\
  2800 Kgs. Lyngby, Denmark\\
  \texttt{climaz@dtu.dk} \\
   \And
 João Lourenço Cardoso \\
  Department of Transportation\\
  National Laboratory for Civil Engineering\\
  101 Av. Do Brasil\\
  1700-066 Lisbon, Portugal \\
  \texttt{jpcardoso@lnec.pt} \\
 \And
 Moshe E. Ben-Akiva \\
  Department of Civil and Environmental Engineering\\
  Massachusetts Institute of Technology\\
  77 Mass Ave, Cambridge, MA 02139\\
  United States of America\\
  \texttt{mba@mit.edu} \\
}

\begin{document}
\maketitle

\begin{abstract}
Traffic microscopic simulation applications are a common tool in road transportation analysis and several attempts to perform road safety assessments have recently been carried out. However, these approaches often ignore causal relationships between different levels of vehicle interactions and/or accident types and they lack a physical representation of the accident phenomena itself. In this paper, a new generic probabilistic safety assessment framework for traffic microscopic simulation tools is proposed. The probability of a specific accident occurring is estimated by an accident propensity function that consists of a deterministic safety score component and a random component. The formulation of the safety score depends on the type of occurrence, on detailed vehicle interactions and maneuvers and on its representation in a  simulation environment. This generic model is applied to the case of an urban motorway and specified to four types of  outcomes: non-accident events and three types of accidents in a nested structure: rear-end, lane-changing, and run-off-road accidents. The model was estimated and validated using simulated microscopic data. To obtained the consistent simulated data, a two-step simulation calibration procedure was adopted: (1) using real trajectories collected on site for detailed behavior representation; and (2) using aggregate data from each event used in safety model estimation.
The final estimated safety model is able to identify and interpret several simulated vehicle interactions. The fact that these outcomes were extracted from simulated analysis shows the real potential of calibrated traffic microscopic simulation for detailed safety assessments.
\end{abstract}


\section{Introduction}
Traffic microscopic simulation applications are a common tool among transportation practitioners and researchers. They enable advanced network efficiency assessment and are able to assess vehicular communication network efficiency, fuel consumption and emissions. Despite several enhancements at the driving behaviour modelling level (for a detailed review see \citet{Barcelo2010a}), the simulation of safety modelling has been neglected due to limitations in the formulations of driver’s perception, decision and error simulation. With the development of many infrastructure-based Intelligent Transportation Systems, research efforts are now identifying traffic conditions that can be statistically linked to accident events. These models are referred to as (real-time) accident probability models; they typically make use of aggregated real-time traffic data collected by sensing technologies (generally from cameras and loop detectors), road geometric characteristics and weather conditions to statistically predict changes in accident occurrence probability. Targeting the assessment of different safety-related solutions, researchers have used these accident probability models to perform safety assessment in microscopic simulation experiments: \citet{Abdel-aty2005} tested the effects of variable speed limits on Interstate 4 in Florida using a matched case-control logistic regression for accident likelihood prediction. Later, the authors developed a nested logit model \cite{Abdel-aty2007a} and an artificial neural network (ANN) accident probability model \cite{Abdel-aty2010a} for predicting accident occurrence using aggregated data from road sensors. These modelling streams rely on the availability of historical accident records and require some level of aggregation of the traffic operations data used as explanatory variables.

Since accidents are considered rare events and it is difficult to isolate the individual effect of the different factors affecting its occurrence, conflicts have been used as an alternative estimator of system safety \cite{Hyden1987}. The use of conflicts has been based on the assumption that the expected number of accidents occurring in a system is proportional to the number of conflicts making suitable for systems’ comparisons \cite{Cunto2008}. One key limitation in using conflicts is how to correctly estimate this proportion. This limitation has motivated the research community to develop several models to estimate accident frequency from traffic conflict frequency \cite{FHWA2008, Wu2012}. Another difficulty in using conflicts for modelling purposes is the lack of practical definitions and measurement standards (as it does not estimate directly the probability of an accident itself). Consequently, several time-based, deceleration-based and dynamic-based surrogate safety performance indicators were proposed in the literature \cite{Laureshyn2010}. These models are most widely used within microscopic simulation studies (see for example \cite{Archer2005,Ozbay2008,Dijkstra2010,Okamura2011,Huang2013}).

Recently, efforts have been made to integrate interaction in probabilistic modelling frameworks. While the aforementioned models try to link the probability of a specific accident occurrence using a statistical model fitted to aggregated data, probabilistic frameworks try to formally represent cause-effect relationships between performed driving tasks and traffic scenarios that may lead to a specific accident event. This approach has better potential to replicate the intrinsic nature of an accident mechanism and, ultimately, does not need to depend on the availability of safety records. On the other hand, these probabilistic frameworks rely on much more detailed information to characterize driving manoeuvres. \citet{Songchitruksa2006} proposed an Extreme Value approach to build relationships between the occurrence of right-angle accidents at urban intersections and frequency of traffic conflicts measured by using the post-encroachment time as accident proximity variable. \citet{Saunier2008} developed a comprehensive probabilistic framework for automated road safety analysis based on motion prediction. For a given interaction between two vehicles, possible trajectories are estimated in a probabilistic framework and the collision probability for a given interaction between two road users can be computed at a given instant by summing the collision probability over all possible motions that lead to a collision. \citet{Wang2010} propose an incident tree model and an incident tree analysis method for the identification of potential characteristics of accident occurrence in a quantified risk assessment framework. These methods provide a more comprehensive formulation of the accident phenomena, but they have not been widely validated or integrated in simulation tools for practical application.

Simulation-based safety studies were also documented in a comprehensive review by \citet{Young2014}. The authors pointed out the need to analyze the probabilistic nature of the link between the vehicle interaction and the accident itself and to generalize the models to accommodate different types of accidents. Furthermore, they recognize the need to differentiate distinct cause-effect relationships for diverse types of accidents and for a probabilistic formulation without the limitations resulting from the aggregation of both traffic data and safety records.

\section{Generic Model Formulation}
\label{sec:generalformulation}

A generic framework for modelling cause-effect mechanisms between detailed vehicle interactions from simulated environments and accident occurrence probability is proposed. It is first assumed that the state of a vehicle $n$ at any given time $t$ can be viewed as a discrete variable whose state outcome $k$ can be one of different types of accident or no accident at all. An individual outcome $k$ among all possible outcomes $K$ is considered to be predicted if its probability $P_{n,t}(k)$ is maximum. As in previous research studies, the main difficulty is how to estimate $P_{n,t}(k)$. This probability should be a function of specific observed variables characterizing the interaction between vehicles and the environment \cite{Tarko2012}. Such premise steps away from the assumption of a fixed coefficient model converting the surrogate event frequency into accident frequency, typically used in the traffic conflicts technique. 

In our proposed framework, the probability for a specific accident involving vehicle $n$ to occur at time $t$ is assumed to be estimable by a specific \textbf{accident propensity} (or proximity) measure \cite{Tarko2009}:

\begin{equation}
P_{n,t}(k)\sim U_{k}
\end{equation}

In the proposed model, each accident propensity function $U_k$ is considered to have a (deterministic) \textbf{safety score} ($V_k$) component and a \textbf{random component} ($\varepsilon$):

\begin{equation}
U_{k}=V_{k}\left(X,\beta\right)+\varepsilon
\label{eq:propensity}
\end{equation}

where $X$ is the vector of explanatory variables, $\beta$ is the vector of unknown parameters to be estimated and $\varepsilon$ is the random term (the terms $n$ and $t$ were omitted here for simplicity). The assumption of a deterministic safety score component aligns with recent research where detailed interaction variables directly affect the accident occurrence probability itself \cite{Wu2012}. The random component $\varepsilon$ is assumed to represent the unobserved effects involved in the determination of the outcome; these may be derived from a random process in the occurrence of a specific event and/or caused by the lack of information from the modelling process.

To formally link the accident phenomenon with the many  variables of interest, the safety score can be formulated as:

\begin{equation}
    V_{k}\left(n,t\right)=f_{k}\left(X_{n,t},X_{n',t},X_{D,t},X_{S}\right)
    \label{eq:systematicUtility}
\end{equation}

where the $k$-accident-type specific scoring function $f_k$ depends on: $X_{n,t}$, the driver-vehicle unit  $n$ specific variables at time $t$; $X_{n',t}$, the variables at time $t$ for the interaction between $n$ and any conflicting driver-vehicle unit $n'$; $X_{D,t}$, the dynamic environmental variables at time $t$ (e.g.: weather, variable speed limit, lighting conditions, etc); and $X_{S}$, the static environmental variables (e.g.: geometrics, road signs, etc). Only a limited set of these variables are modelled in microscopic traffic simulation and, therefore, available for integration in safety models for simulated environments. Furthermore, different simulators may consider different sets of variables (e.g.: freeway simulator vs. urban road network simulator), which further limits the number of available candidate explanatory variables $X$.
In the presented model we framed the generic formulation: 

\begin{itemize}
    \item to represent cause-effect relationships by means of each function $f_k$; 
    \item to rely on surrogate safety measures as potential descriptors of the interaction variables $X_{n',t}$; 
    \item to simultaneously deal with different non-independent types of accident outcomes; 
    \item to consider a disaggregated probability for any vehicle state ($n,t$) observation instead of the existing aggregate formulation used in state-of-the-art real-time accident probability models.
\end{itemize}

\section{Modelling Different Accident Types}

Research in accident prediction and accident frequency modelling (see \cite{Lord2010}, for a comprehensive review) has led to the development of advanced statistical capabilities for modelling the relationship between different road and traffic characteristics with different accident types. However, this work has been limited to the available data, which is typically in aggregated form and restricts further progress in the related findings. Although some of the existing findings may help in the specification of the safety score formulation for different accident types (such as the non-linear relationship between speed and run-off-road accidents), the formal link between a set of very specific disaggregated variables, typically available in a simulated environment and accident probability (e.g.: available side gaps, relative speed and lane changing conflicts) has not been specified. The existing correlations between the available artificial variables should also be considered. Furthermore, the dependency of the formulation itself on the available simulated variables further increases the variability of the modelling specification task. Thus, in the present paper, the above general formulation is only detailed for a specific set of accidents that typically occur on busy urban motorways and under the linearity framework: rear-end accidents (RE), side collisions during lane-change maneuvers (LC) and run-off-road accidents (ROR), i.e. \textit{K}=\{RE, LC, ROR\}. These three different outcomes correspond to very distinct phenomena. However, these three outcomes may be related, namely if one considers accident outcomes following an evasive action from different risky interactions. Thus, the general model assumes the following nested structure (see Figure \ref{fig:modelstruct}):

\begin{figure}[!htb]
  \centering
  \includegraphics[width=0.50\textwidth]{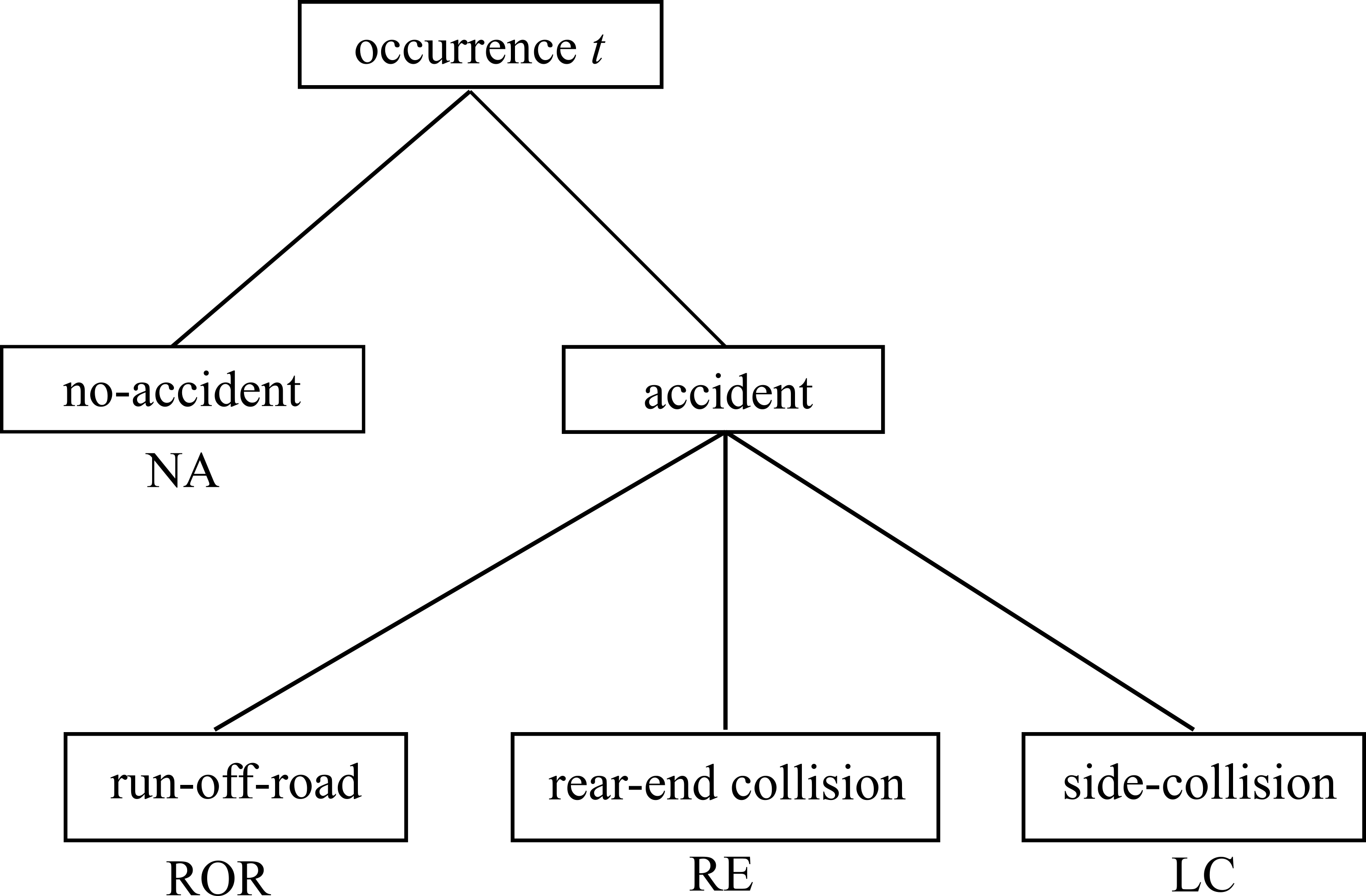}
  \caption{Model structure for motorway accident occurrence.}
  \label{fig:modelstruct}
\end{figure}

When selecting the explanatory variables for each accident specific safety score, $V_k(n,t)$, (eq. \ref{eq:systematicUtility}), we face a limitation in the available simulated variables. The safety scores of the three types of accidents presented in the next sections rely on typical variables available in microscopic traffic simulation tools. Yet, the generic proposed model is flexible enough to allow the integration of more detailed simulated variables (e.g.: steering angle). The linearity assumption is considered due to the absence of a readily available optimized functional-from using several detailed motion variables. These complex relationships deserve a more thorough investigation and are not analyzed in the current paper.

\subsection{Rear-end (RE) conflicts}
\begin{figure}[!htb]
  \centering
  \includegraphics[width=0.45\textwidth]{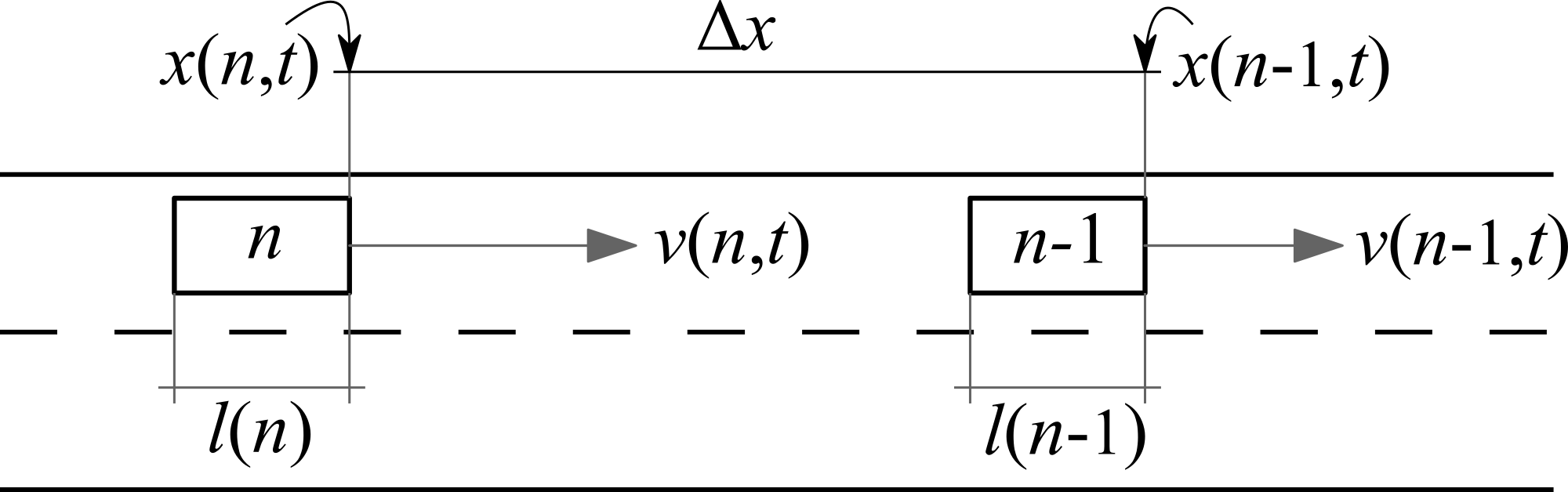}
  \caption{Rear-end interaction.}
  \label{fig:REdiag}
\end{figure}

When facing rear-end interactions the probability of a collision is formulated in terms of the subject vehicle braking requirements to avoid a RE collision and the maximum available braking power. The first,, is represented by the difference between the actual relative acceleration between the subject vehicle and its leader, $\Delta a^{need}$, and the \textbf{D}eceleration \textbf{R}ate required to \textbf{A}void \textbf{C}rash (\textit{DRAC}), estimated using Newtonian physics.

\begin{equation}
\Delta a^{need}\left(n,t\right)=DRAC\left(n,t\right)+\Delta a\left(n,t\right)\label{eq:deltaA}
\end{equation}

where $\Delta a = a(n,t)-a(n-1,t)$ is the acceleration difference between the vehicle $n$ and its leader $n-1$, and

\begin{equation}
DRAC\left(n,t\right)=\frac{\left[v\left(n,t\right)-v\left(n-1,t\right)\right]^{2}}{2\left[x\left(n-1,t\right)-x\left(n,t\right)-l\left(n-1\right)\right]}\label{eq:drac}
\end{equation}

where $\Delta a^{need}\left(n,t\right)$ is the needed deceleration to reach the \textit{DRAC} for the subject vehicle $n$ at time $t$, $\Delta a\left(n,t\right)=a\left(n,t\right)-a\left(n-1,t\right)$ is the acceleration difference between the subject vehicle and its leader, and $v\left(n,t\right)$, $x\left(n,t\right)$ and  $l(n)$ are the speed, longitudinal (front bumper) position and length of vehicle $n$ (see Figure \ref{fig:REdiag}).

We further split $\Delta a^{need}$ into its positive ($\Delta a^{need}_{+}(n,t)>=0$) and negative  ($\Delta a^{need}_{-}(n,t)<=0$) components, thus allowing for the consideration of different model parameters for different safety scenarios, i.e. when the relative acceleration is larger or smaller than the \textit{DRAC}. The advantage of using $\Delta a^{need}$ instead of just the \textit{DRAC} is the consideration of the current acceleration state. The value of $\Delta a^{need}$ is easily interpreted: the negative values represent safer values, for which the vehicle relative deceleration is greater than \textit{DRAC}.
We further improve this formulation by dividing the required deceleration by the time-to-collision (\textit{TTC}), thus considering how long the driver has before the potential collision. The safety score function will then depend on the available time for adjustment, resulting in a relative needed deceleration ratio, $RA^{need}(n,t)$:

\begin{equation}
\begin{dcases} 
RA_{+}^{need}\left(n,t\right)=\frac{\Delta a_{+}^{need}\left(n,t\right)}{TTC\left(n,t\right)}\\
RA_{-}^{need}\left(n,t\right)=\frac{\Delta a_{-}^{need}\left(n,t\right)}{TTC\left(n,t\right)}
  \end{dcases}
\label{eq:RA}
\end{equation}

\begin{equation}
TTC\left(n,t\right)=\frac{x\left(n-1,t\right)-x\left(n,t\right)-l\left(n-1\right)}{\left(v\left(n,t\right)-v\left(n-1,t\right)\right)}
\label{eq:ttc}
\end{equation}

Finally, similarly to the crash potential index (a surrogate safety measure described in \cite{Cunto2008}, a measure of the excess regarding the maximum available deceleration rate is also considered. It considers heterogeneous safety conditions regarding different vehicle categories and different pavement conditions (e.g.: dry/wet) that are expected to influence the deceleration performance:

\begin{equation}
\Delta a^{lim}\left(n,t\right)=DRAC\left(n,t\right)-\left(\mu^{long}\left(n,t\right)+d\right)g
\label{eq:deltalimit}
\end{equation}
\begin{equation}
\mu^{long}\left(n,t\right)=f^{long}\left(v\left(n,t\right),\alpha^{type},\alpha^{surf}\right)
\label{eq:mulong}
\end{equation}

where $\Delta a^{lim}$ is the excess to the maximum available deceleration for vehicle $n$ at time $t$, $DRAC\left(n,t\right)$ is the deceleration rate required to avoid crash presented above, $d$ is the grade rate (m/m), $g$ is the gravitational acceleration of 9.81 m/s$^{2}$ and $\mu^{long}(n,t)$ is the maximum available longitudinal friction coefficient, which depends on the speed of the vehicle itself $v(n,t)$ and on two factors that account for the vehicle type, $\alpha^{type}$, and the pavement condition, $\alpha^{surf}$. This simplified formulation of the friction coefficient is due to the small number of variables typically available in simulated environments.

Similar to the previous relative needed deceleration ratio, $RA^{need}(n,t)$, we compute the  rate $RA^{lim}(n,t)$ as the physical measure of the excess to the allowable jerk to account for the available \textit{TTC}:

\begin{equation}
RA^{lim}(n,t)=\frac{\Delta a^{lim}\left(n,t\right)}{TTC(n,t)}
\end{equation}

Finally, the systematic safety score for RE collisions is formulated as:

\begin{equation}
V_{RE}\left(n,t\right)=\beta_{0}^{RE}+\beta_{1}^{RE}RA_{+}^{need}\left(n,t\right)+\beta_{2}^{RE} RA_{-}^{need}\left(n,t\right)+\beta_{3}^{RE} RA^{lim}\left(n,t\right)
\label{eq:REscore}
\end{equation}

where $RA_{+}^{need}$ and $RA_{+}^{need}$ are the positive and negative components of the relative needed deceleration ratio computed using $\Delta a_{+}^{need}$ and $\Delta a_{-}^{need}$ respectively; $RA^{lim}$ is the excess to the allowable jerk; and $\beta_{0}^{RE}$, $\beta_{1}^{RE}$, $\beta_{2}^{RE}$ and $\beta_{3}^{RE}$ are the estimable parameters.

\subsection{Lane-change (LC) conflicts}

\begin{figure}[!htb]
  \centering
  \includegraphics[width=0.45\textwidth]{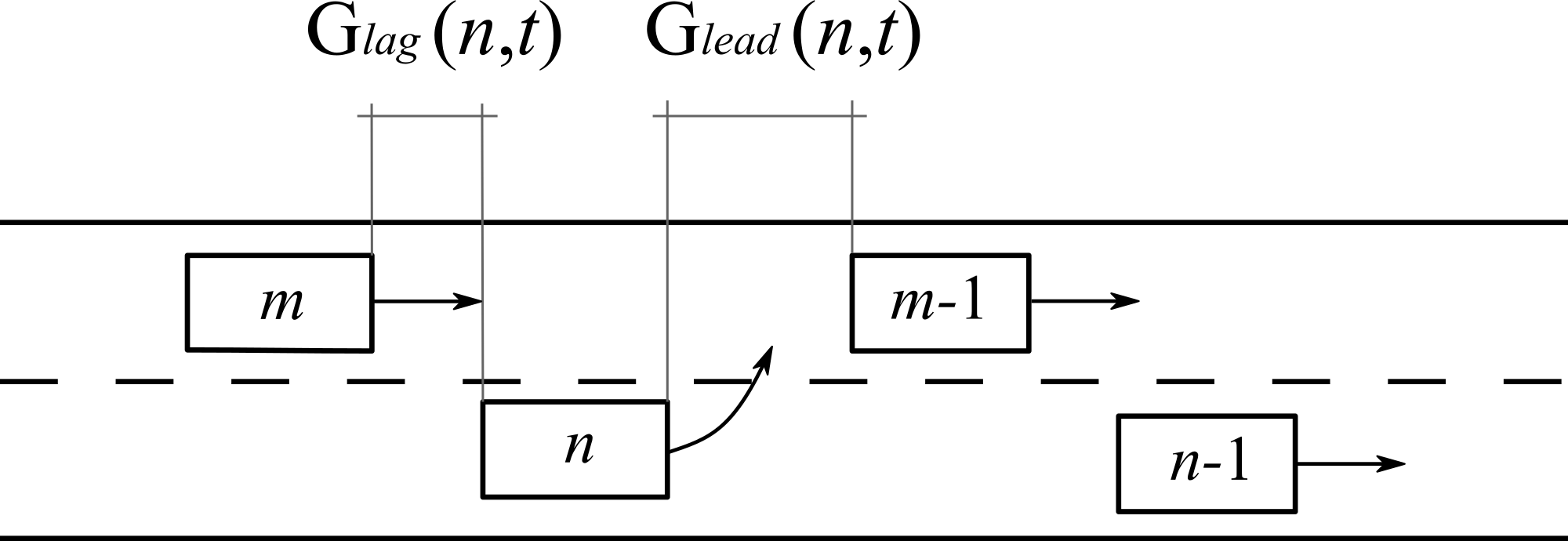}
  \caption{Lane-change interaction.}
  \label{fig:LCdiag}
\end{figure}

The lane change action decision is typically modelled by means of gap acceptance models \cite{Toledo2007} or by acceleration variation models \cite{Kesting2007}. One would expect that the probability of lane-change collisions also depends on vehicle lateral movements. However, the majority of current micro-simulation tools do not integrate a detailed modelling of the lane-changing manoeuvre or rely on external tools for lane-change trajectory reconstruction \cite{So2015}. Therefore, surrogate measures depending on lateral movements, such as the time-to-lane-crossing proposed by \citet{VanWinsum1999} and the post-encroachment-time used by \citet{Zheng2013} are not easily integrated. To complicate further, gap acceptance is generally modelled separately for the lead and lag gaps on the target lane (see Figure \ref{fig:LCdiag}) \cite{Toledo2007}. 

This disaggregation is of special interest as different parameters may be computed for different gaps \cite{Chovan1994}. The lane-changing process becomes increasingly difficult as the speed differences between the subject vehicle and the lead and lag vehicles in the target lane increases \cite{Hidas2005a}. Thus, in the proposed formulation, the safety score of the  event is specified in terms of relative gap variation:

\begin{equation}
RG^{gap}(n,t)=\frac{\Delta v^{gap}(n,t)}{G^{gap}(n,t)}
\label{eq:RGap}
\end{equation}

where $G^{gap}$ is the gap in meters and $\Delta v^{gap}$ represents the speed difference in m/s between the subject vehicle and the lead or lag vehicle on the target lane, with $gap=\{leand,lag\}$:

\begin{equation}
\Delta v_{tl}^{lead}(n,t)=v(m-1,t)-v(n,t)
\label{eq:leadspeed}
\end{equation}
\begin{equation}
\Delta v_{tl}^{lag}(n,t)=v(n,t)-v(m,t)
\label{eq:lagspeed}
\end{equation}

where $v(m-1,t)$ and $v(m,t)$ are the speed of the lead vehicle $m-1$ or the lag vehicle $m$ in the target lane, respectively. Again, the split of the relative gap variation into its positive, $RG^{gap}_{+}$, and negative, $RG^{gap}_{-}$, values allows for the consideration of different parameters associated with different safety conditions, i.e. for gaps that are either increasing or decreasing, respectively.

\begin{equation}
\begin{dcases}
RG_{+}^{gap}(n,t)=\max(0,R^{gap}(n,t))\rightarrow RG_{+}^{gap}(n,t)\geq0\\
RG_{-}^{gap}(n,t)=\min(0,R^{gap}(n,t))\rightarrow RG_{-}^{gap}(n,t)\leq0
\end{dcases}
\label{eq:twoRGaps}
\end{equation}

with $gap=\{lead,lag\}$. Following the above formulation a gap with a higher relative shrinking rate ($RG_{-}^{gap}(n_1,t_1)< RG_{-}^{gap}(n_2,t_2)\leq0$) should have a higher impact on the  LC conflict probability, $P_{n_1,t_1}(LC)>P_{n_2,t_2}(LC)$ and thus, its parameter estimate should be negative.
The systematic component for LC collisions may now be formulate as: 

\begin{equation}
V_{LC}(n,t)=\beta_{0}^{LC}+\beta_{1}^{LC}RG_{+}^{lag}(n,t)+\beta_{2}^{LC}RG_{-}^{lag}v+\beta_{3}^{LC}RG_{+}^{lead}(n,t)+\beta_{4}^{LC}RG_{-}^{lead}(n,t)
\label{eq:LCscore}
\end{equation}

where $RG_{+}^{lag}$, $RG_{-}^{lag}$, $RG_{+}^{lead}$ and $RG_{-}^{lead}$ are the positive and negative components of the lag and lead gaps, respectively; and $\beta_{0}^{LC}$, $\beta_{1}^{LC}$, $\beta_{2}^{LC}$, $\beta_{3}^{LC}$ and $\beta_{4}^{LC}$ are the estimable parameters.

\subsection{Run-off-road (ROR) events}
ROR events are assumed to be primarily related to individual vehicle dynamics rather than interaction related variables. This assumption is especially true under free-flow scenarios. However, it may also result from evasive manoeuvres due to risky lane-changing or car-following decisions.

Vehicle dynamics in traffic simulation models are represented in a much simplified manner when compared with the detailed movements’ description of real events and its representation currently achieved with accident reconstruction models. This significantly limits the current potential for a ROR micro-simulation modelling framework. The vehicle lateral movement, the true road geometric characteristics (such as transition curves), the pavement surface characteristics, and the vehicle detailed physical and mechanical attributes are generally not available. However, some relevant variables that may potentially be useful for the analysis of ROR events are already available in micro-simulation tools, such as vehicle speed, general road geometrics and the generic vehicle type.

\begin{figure}[!htb]
  \centering
  \includegraphics[width=0.45\textwidth]{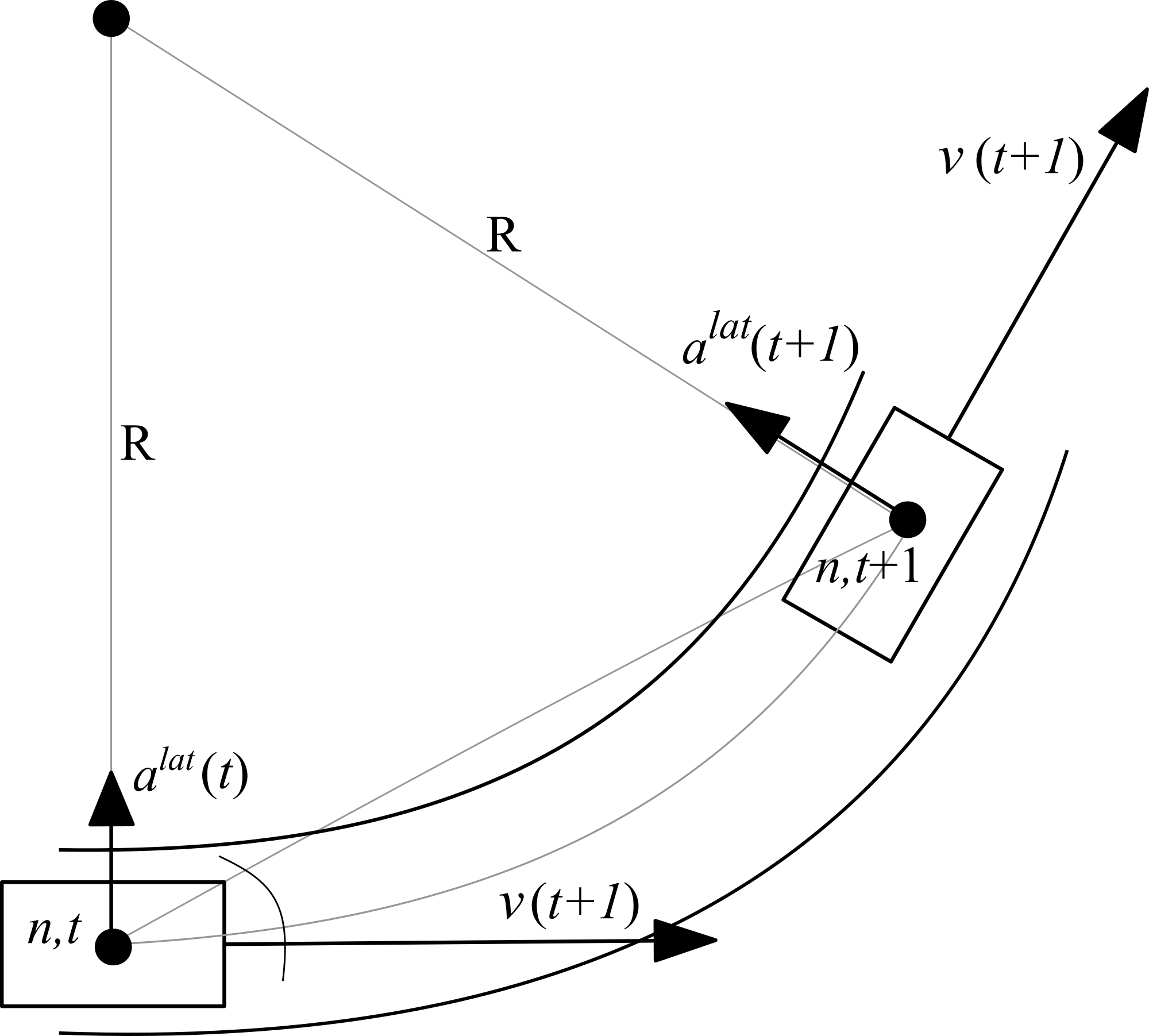}
  \caption{Run-off-road event.}
  \label{fig:RORdiag}
\end{figure}

In the proposed framework, the safety score of ROR events is assumed to be linked to the difference between the current lateral acceleration of vehicle $n$, $a^{lat}(n,t)$, and a site specific critical lateral acceleration, $a^{lat}_{cr}(n,t)$. Since vehicle lateral movements are not modelled, the vehicle path on curve elements is assumed to be a simple circular path and the vehicle yaw equal to the curve bearing (see Figure \ref{fig:RORdiag}).

The lateral acceleration of vehicle $n$, $a^{lat}$, is derived from its current speed and the curve radius $R$ in meters:

\begin{equation}
a^{lat}(n,t)=\frac{\left[v(n,t)\right]^2}{R}
\label{eq:aLat1}
\end{equation}

Although the majority of the simulation tools do not provide information on lateral movement during a lane change, we still expect that this event \textit{per se} can be included in the ROR safety score function. Using test track data, \citet{Chovan1994} considered peak lateral acceleration values of 0.4${g}$, 0.55${g}$ and 0.7${g}$ for mild, moderate, and aggressive steering manoeuvres, respectively. As detailed lane change models such as the one formulated by \citet{Chovan1994} are typically not available in microscopic traffic simulation platforms, a generic peak acceleration add-on for lane change of 0.5${g}$ was adopted and integrated in eq. \ref{eq:aLatfull} to account for a potential increased probability in both curved and straight road sections during a lane change:

\begin{equation}
a^{lat}(n,t)=\frac{\left[v(n,t)\right]^2}{R}+0.5g \delta^{lc}(n,t)
\label{eq:aLatfull}
\end{equation}

where $\delta^{lc}(n,t)$ is a dummy variable to account for lane change (1 if the vehicle is performing a lane change, 0 otherwise).

The maximum allowed lateral acceleration $a^{lat}_{cr}(n,t)$ directly depends on the critical lateral friction coefficient $\mu^{lat}$ and the road super-elevation $e$ (m/m):

\begin{equation}
a^{lat}_{cr}(n,t)=(\mu^{lat}(n,t)+e)g
\label{eq:acritical}
\end{equation}

Similar to its longitudinal counterpart, the values of the maximum lateral friction coefficient, $\mu^{lat}$, also depend on the vehicle speed itself $v(n,t)$, on the pavement condition and on the type of vehicle:

\begin{equation}
\mu^{lat}(n,t)=f^{lat}(v(n,t),\alpha^{type},\alpha^{surf})
\label{eq:mulat}
\end{equation}

The safety score function may now be formulated in terms of the positive (unsafe) and negative (safe) components of the difference between the current and the critical lateral accelerations:

\begin{equation}
V_{ROR}(n,t)=\beta_{0}^{ROR}+\beta_{1}^{ROR}\Delta a_{+}^{lat}(n,t)+\beta_{2}^{ROR}\Delta a_{-}^{lat}(n,t)
\label{eq:RORscore}
\end{equation}

where $\Delta a_{+}^{lat}$ and $\Delta a_{-}^{lat}$ are the positive and negative components of $\Delta a^{lat} = a^{lat}-a^{lat}_{cr}$, respectively.

\subsection{Estimation framework}
As previously stated, the explanatory variables of one type of accident may influence the occurrence of others and evasive manoeuvres may create correlations between different accident outcomes. When modelling multiple discrete outcomes, the multinomial nested logit model (NL) proposed by \citet{BenAkivaPhD} has advantages over the simple multinomial logit model. It can simultaneously estimate the influence of independent variables while allowing for the error terms to be correlated and, therefore, allowing for the violation of the independence of irrelevant alternatives (IIA) property. The general NL is therefore here assumed, using the safety scores specified in eq. \ref{eq:REscore}, \ref{eq:LCscore} and \ref{eq:RORscore} as the systematic component of the alternatives of rear-end (RE), lane-changing (LC) and run-off-road (ROR) accident outcomes, respectively. The no-accident event (NA) was defined by an alternative specific constant alone and the NL was normalized at the top (the reader is referred to reference books such as \cite{BenAkivaBook}, for further details of the NL model).

To directly estimate the proposed model, a large set of events with all types of model safety outcomes and the observed vehicle interactions is needed. Unfortunately, a large data set with real trajectories for several observed events (accident and non-accident) is still not available. Hopefully, data sets such as the Naturalistic Driving Study of the Strategic Highway Research Program 2 will fill this gap in the near future \cite{McLaughlin2015} and will form a comprehensive data set for estimation of the proposed model. Furthermore, as mentioned in section \ref{sec:generalformulation}, we conditioned the formulation of the proposed framework to the limitations of simulation tools for integration purposes. Thus, we rely on the use of simulated (artificial) trajectories generated for each of the available recorded safety events in the estimation process. Yet, a set of critical assumptions must be considered:

\begin{enumerate}

\item The philosophy of microscopic simulation applications is to replicate real aggregate measurements as closely as possible, including detailed variables as accelerations, headways or \textit{TTC}. We assume that a well calibrated microscopic simulation model is able to replicate the statistical distributions of detailed traffic variables. For that, we also expect that the microscopic model will be calibrated appropriately using measurements for such detailed data. The importance of using disaggregated data such as trajectories in the analysis of detailed driving behaviour and vehicle interactions has been pointed out in several studies \cite{Cunto2008, LimaAzevedo2015}. 

\item Trajectories extracted in a generic day represent the general driving behaviour of traffic. This assumptions ensures that a calibrated of simulator under assumption 1 above will produce distribution of detailed trajectory variables representative of the realistic conditions on site. Confidence in this assumption depends on the amount and breath of information available. However, other factors (such as weather) are expected to influence the driving behaviour model parameters. Thus, we assumed that part of this variability will be accounted using a second dedicated calibration, carried out for each specific observed event considered in estimation and using aggregated data sets, such as sensor data.

\item A link between detailed traffic variables and accident occurrence is assumed. Even if simulation models are accident free, it is assumed that the description of detailed traffic variables can be linked to the accident probability and, therefore, that traffic characteristics affect safety. This assumption is supported by several previous studies (e.g.: \cite{Archer2005, Abdel-aty2007a}).

\end{enumerate}

Based on the above assumptions, the microscopic simulation tool that will generate the artificial data for estimation needs to be calibrated for each of the observed/recorded events. As mentioned in assumption 1, the use of real trajectories is essential. In order to appropriately generate artificial data for the model estimation, a two-step procedure is used. First, the simulator needs to be calibrated once using the disaggregated (observed vehicle trajectories) data collected for a specific generic day; the purpose of the first calibration is to tune the full set of behavioural model parameters to the generic conditions observed in the network. This calibrated parameters set is then used as the initial set in subsequent (fine-tuning) calibrations for each event observation  used in the safety model estimation and using the event-specific aggregate (loop sensor-based) data available. This method was proposed and validated in \cite{LimaAzevedo2015}; the optimum set of parameters calibrated for each event $i$ is used to generate the artificial detailed traffic variables for that same event event $i$ alone. Finally, the full set of detailed artificial traffic variables is used jointly with its associated outcome $k$ for event $i$ (RE, LC, ROR and NA) to estimate the proposed safety model.

The proposed model is specified individually for any vehicle $n$ at every time $t$. However, the use of artificial trajectories does not allow us to directly link a specific trajectory with the event outcome. Furthermore, it is typical that both the loop-based variables used for the event specific calibration and the variables collected in accident reports are defined for pre-defined time and spatial units. In some cases, such aggregated intervals may be too large to capture short-term variations; nevertheless several authors \cite{Oh2001,Abdel-aty2005}) have successfully used aggregated periods of up to 5 minute intervals to perform accident occurrence probability analyses. With the absence of the true trajectories for vehicle $n$ involved in each observed event $i$, the characterization of the detailed traffic variables for a specific accident occurrence must be linked by means of spatial and temporal aggregation. Therefore, the estimation needs to aggregate all vehicle state outcome probabilities $P_{n,t}(k)$ by standardized intervals of space $s$ and time period $p$:

\begin{equation}
P_{s,p}(k)=\frac{1}{N} \sum_{N} P_{n,t}(k)
\label{eq:DiscreteProbability}
\end{equation}

here $P_{n,t}(k)$ is the probability of occurrence $k$ for any relevant observation of vehicle $n$ at time $t$, traveling in spatial interval $s$ and time period $p$ and defined by the proposed NL model; $P_{s,p}(k)$ is the probability of occurrence $k$ for the specific spatial interval $s$ and time period $p$. $N$ is the total number of observations for all vehicles that travelled in the interval $s,p$. It is important to point out that the final estimated model is still based on individual $n,t$ probabilities. However, the model estimation is based on the distribution of general microscopic traffic variables and not on extreme values alone. This follows the general traffic micro-simulation philosophy where the replication of averaged variables is expected. While it is possible to explore the use of extreme value targeted formulations, specific calibration methods and potential extensions of driver behaviour model specifications for near-misses should be considered and are beyond the scope of the present research.
Finally, if one considers a large observation period that is typically needed to have a sufficient number of accident occurrences, it is expected that the loop sensors will fail in some instances. The computational memory and processing resources needed to generate and use the simulated trajectory data for a large set of no-accident occurrence units is impractical. For this purpose outcome (choice)-based random sampling was assumed. Then, to account for this biased sampling process, the weighted exogenous sample maximum likelihood function (WESML) proposed by \citet{Manski1977} was used.

\section{The Urban Motorway Case}

\subsection{Network and data collection}
\begin{figure}[!htb]
  \centering
  \includegraphics[width=0.60\textwidth]{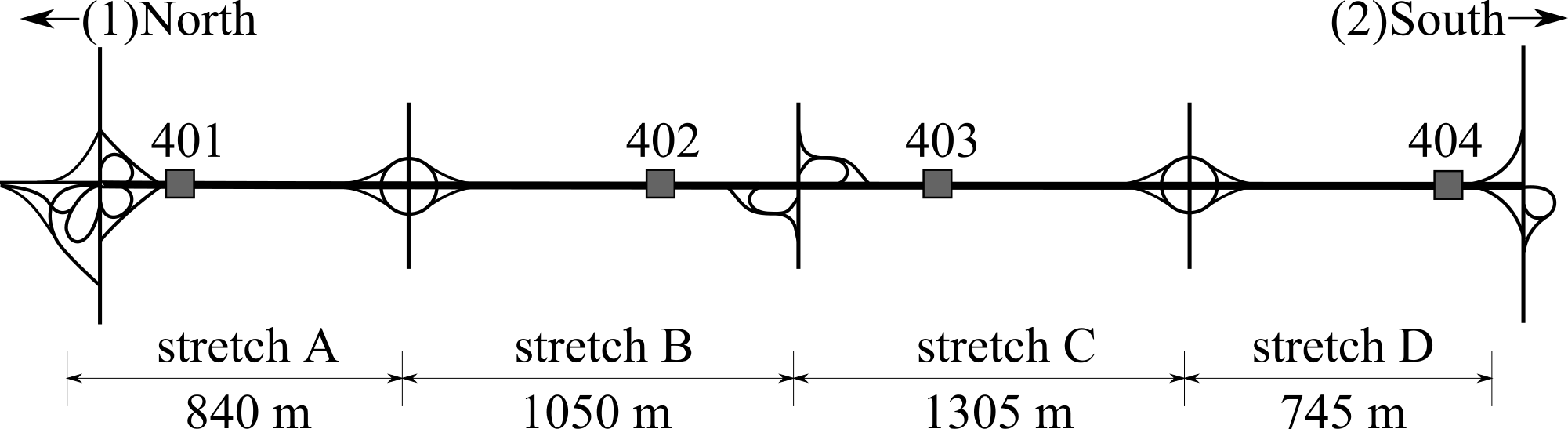}
  \caption{A44 layout.}
  \label{fig:A44sketch}
\end{figure}

The proposed model was estimated using collected and simulated data for the A44 urban motorway near Porto, Portugal. This road was selected as a case study due to its dense traffic during peak hours, unusually high number of lane changes, short spacing between interchanges and high percentage of heavy goods vehicles. A44 is a 3,940m long dual carriageway urban motorway with 5 major interchanges, two 3.50m wide lanes and 2.00m wide shoulders in each direction (see Figure \ref{fig:A44sketch}). There are acceleration and deceleration lanes at all interchanges, and several are as short as 150m. On and off-ramps connect to local roads which generally have tight horizontal curves, intersections or pedestrian crossings- features that impose significant reductions in vehicle speeds.

Three different traffic data sets were specifically collected for the present study:
\begin{itemize}
\item A dynamic seed Origin-Destination matrix (OD) based on a sample of license plate matching and vehicle counts \cite{LimaAzevedo2014};
\item 5 min loop sensor speeds and counts for the existing eight traffic stations (4 in each direction: 401, 402, 403 and 404 in Figure \ref{fig:A44sketch}), between 2007 and 2009. This resolution results in a total of 315,360 observations for each station. After removing the erroneous records, the daily statistics algorithm (DSA) proposed by \citet{Chen2003} was used. This algorithm uses the time series of flow and occupancy measurements to detect abnormal values, instead of signalling data problems just based on an individual observation (the details on the application of the DSA to the present case study can be found in \cite{LimaAzevedo2014});
\item Vehicle trajectories collected for a generic morning by aerial remote sensing for the entire length and access links of the A44 motorway and for congested and non-congested periods \cite{LimaAzevedo2014}. A light aircraft overflew the A44, between 8:45 and 10:45 AM on 11\textsuperscript{th} October 2011. Flight characteristics were optimized considering the atmospheric conditions on-site and a specific desired ground sample distance. Images were orthorrectified using a 3D terrain model, with the camera and lens characteristics and the precise flight positioning data recorded through differential GPS. Images were collected at an average rate of 0.5Hz. The vehicle detection on each collected image was carried out using coloured background subtraction. The trajectory was then reconstructed using an extension of the k-shortest disjoint path algorithm using motion-based optimization. A total of 1855 partial trajectories were successfully collected. For further details on the trajectory extraction method and its validation the reader is referred to \cite{LimaAzevedo2014}. Finally, aggregated speed and count data (sensor based) for this data collection period was compared and validated against the observations for 2007 to 2009 for conditions stability.
\end{itemize}

Incident records were also collected for the same period of 2007 to 2009 including a total of 144 rear-end collisions (44\%), side-collisions during lane-changing manoeuvres (24\%) and run-off-road accidents in curved segments or during lane-changing manoeuvres (32\%). One fatality, 7 serious injuries and 98 slight injuries were registered in the area during this period. A significant share (33\%) of the accidents was located on the edging sections of the A44 (up to 500m), near the interchanges of stretch A and D. This is due to the motorway layout in these specific areas, with frequent lane changes and speed variations due to the appropriate route choice manoeuvres. Fifty-eight percent of the accidents were located on straight segments which may partially explain the lower share of run-off-road accidents in the sample. All accidents were manually geo-referenced to better pinpoint their occurrence during the simulation experiments.

Along with the 5 min temporal units for the observed traffic data, the nature of the accident location record carried out by the authorities required a spatial observation unit of 50 m. These units are considered for the aggregation of individual probabilities. Using such units, a very large number of no-accident (NA) events were observed during this three years period (more than $180 \times 10^{6}$). After excluding the days with erroneous sensor data, a random sampling technique was used to select 6,400 no-accident events, resulting in a total of 6,544 events to be calibrated and simulated for artificial data generation.

\subsection{Driving behaviour model and simulation tool}

The integrated driver behaviour model presented in \cite{Toledo2007} was used to simulate trajectories for each observed event. It integrates four levels of tactical decision-making (target lane, gap acceptance, target gap and acceleration) in a latent decision framework based on the concepts of short-term goal and short-term plan. A probabilistic model is used to capture drivers’ lane choice decisions and driving behaviour parameters and vehicle characteristics are randomly assigned to each driver-vehicle unit. This model has been integrated in the open-source traffic simulation platform \texttt{MITSIMLab} \cite{Yang2000}. \texttt{MITSIMLab} is a simulation-based laboratory that was initially developed for evaluating the impacts of alternative traffic management system designs at the operational level and assisting in subsequent design refinement. Traffic and network elements are represented in detail in order to capture the sensitivity of traffic flows to the control and routing strategies. \texttt{MITSIMLab} has been successfully applied in several traffic and research studies in the USA, the UK, Sweden, Italy, Switzerland, Japan, South Korea, Singapore and Malaysia. 

\subsection{Calibration method}
For the first disaggregate calibration of the full set of \texttt{MITSIMLab}’s driving behaviour model parameters, the global multi-step sensitivity-analysis based calibration proposed in \cite{Ciuffo2014} was used along with real trajectories. A set of nine performance measures that were used in the calibration were selected to describe the trajectory data: speed, acceleration, deceleration, headway, \textit{TTC}, \textit{DRAC}, number of lane-changes, and lead and lag gaps during a lane-change. For each of these variables (except for number of lane-changes), 11 statistics were used to characterize their distribution for the entire network: the minimum value, nine percentiles (10th, 20th, ..., 90th) and the maximum value in the sample. The goodness-of-fit measure used in the selection of the best parameter combination was the Theil’s inequality coefficient (see \cite{Hollander2008} for its formulation and advantages in the calibration of traffic simulation models). The demand used in the calibration was estimated for the specific-day of the trajectory extraction using the dynamic seed OD and the respective loop count data. A result of this calibration process is presented in Table 1.

\begin{table}[!htb]
 \caption{Theil’s Inequality Coefficient after calibration using trajectories}
  \centering
  \begin{tabular}{lllllllll}
    \toprule
    \cmidrule(r){1-2}
    Speed   & Acceleration  & Deceleration  & Headway	& \textit{TTC}	& \textit{DRAC}	& Lane-changes	& Lead gaps	& Lag gaps  \\
    \midrule
    0.075	& 0.279	& 0.273	& 0.160	& 0.136	& 0.889	& 0.742	& 0.197	& 0.201 \\
    \bottomrule
  \end{tabular}
  \label{tab:theilresults}
\end{table}

A powerful simultaneous demand-supply calibration method \cite{Lu2015} was then used for the event-specific calibration of the accident and non-accident events in record, using the count and speed aggregated data of the 8 loop sensors in the network. At this stage, the parameter set to be calibrated is composed of the dynamic OD pairs of the A44 and a subset of 11 (out of 101) sensitive parameters of the driving behaviour model (identified using global sensitivity analysis, \cite{Ciuffo2014})\footnote{For a complete list of the 101 parameters of the driving behaviour model of \texttt{MITSIMLab}, the reader is referred to \cite{Ciuffo2014}}. The measures of performance used in this stage were loop sensor speed and count measurements. This process was carried out 6,544 times, once for each event in the estimation sample and using a high performance computer for faster computation. The calibration was based on the traffic conditions for the 30 min periods before the occurrence and on the trajectory-based calibrated parameters as starting point. The performance of the WSPSA algorithm was quite satisfactory, with average reductions of the root mean square normalized error to 28.2\% and 27.6\% regarding the speed and count loop sensor data, respectively. For further details on the calibration procedure, the reader is referred to \cite{LimaAzevedo2015}.

\subsection{Artificial data generation}
After the calibration of each of the recorded accident events and sampled non-accident events the artificial data was generated. For the accident occurrences, the 144 simulations resulted in an average of about $1.5\times 10^5$ observations of vehicle motion variables at a frequency of 1Hz. These observations were recorded for the 50 m section upstream the accident location, and within the 5 min period before its occurrence. The 6,400 ramdomly sampled non-accident events resulted in about $4.5 \times 10^6$ observations for the same spatial and temporal units.

Given the number of observations a large number of replications, while desirable when working with simulated data, is unmanageable during the estimation process. Thus, only three replications of each event were performed. However, a sensitivity analysis on the model prediction capabilities regarding the artificial data variability is presented in Section \ref{sec:validation}.

The artificial data generated by the calibrated models showed divergence between the simulated outputs for accident and non-accident events which supports the usefulness of advanced traffic simulation tools in the replication of detailed interactions and driving behaviour (see Table \ref{tab:artificialdata}). The observed differences from the artificial data to the real sample comes from the global calibration based on multiple measures of performance for both disaggregated and aggregated data.

\begin{table}[!htb]
 \caption{Summary statistics of artificial trajectories for the 5 min and 50m before the simulated accident and non-accident events and of real trajectories collected on-site.}
  \centering
  \begin{tabular}{llccc}
    \toprule
    \multicolumn{2}{c}{Variable}    & Mean	& Std. dev.	& Median\\
    \cmidrule(r){1-5}
    \multirow{3}*{Speed ($m/s$)}  & Accidents	& 12.51	& 10.00	& 12.19\\
    & Non-accidents	& 18.97	& 8.78	& 19.50\\
	& Real trajectories	& 21.73	& 7.38	& 22.84\\
    \midrule
    \multirow{3}{*}{Acceleration ($m/s^2$)}  & Accident	& 1.17	& 0.89	& 0.93\\
	& Non-accident	& 0.79	& 0.61	& 0.71\\
	& Real trajectories	& 0.66	& 0.87	& 0.30\\
    \midrule
    \multirow{3}{*}{Deceleration ($m/s^2$)}	& Accident	& -1.10	& 0.92	& -0.87\\
	& Non-accident	& -0.92	& 0.86	& -0.72\\
	& Real trajectories	& -0.85	& 0.90	& -0.44\\
	\midrule
	\multirow{3}{*}{Headway (\textit{m})}	& Accident	& 21.83	& 29.87	& 26.80\\
	& Non-accident	& 38.23	& 34.57	& 29.50\\
	& Real trajectories	& 45.84	& 33.90	& 35.86\\
	\midrule
    \multirow{3}{*}{Lead side gap before a lane change (\textit{m})}	& 	Accident	& 4.49	& 6.95	& 1.90\\
	& Non-accident	& 9.68	& 10.91	& 4.5\\
	& Real trajectories	& 11.90	& 8.87	& 10.53\\
    \midrule
    \multirow{3}{*}{Lag side gap before a lane change (\textit{m})}	& 	Accident	& 3.68  & 5.37	& 1.87\\
	& Non-accident	& 10.19	& 8.71	& 8.56\\
	& Real trajectories	& 12.46	& 8.99	& 11.77\\
    \bottomrule
  \end{tabular}
  \label{tab:artificialdata}
\end{table}

\section{Estimation Results}
\subsection{Modeling assumptions}
For the computation of the RE and ROR model components, both $\mu^{long}$ and $\mu^{lat}$ must be specified (see equations \ref{eq:mulong} and \ref{eq:mulat}. Unfortunately, measured values on-site are not available. Generic $\mu_0$ values were adopted based on measurements from other urban freeways found in the literature, considering the pavement surface factor ($\alpha^{surf}$) function of its dry and wet condition as a direct variation from 0.85 at 0km/h to 0.75 at 130km/h for dry pavements and from 0.70 at 0km/h to 0.20 at 130km/h for wet pavements \cite{Inoue1993}. An increase factor of 1.10 was considered for the lateral coefficient $\mu^{lat}$. Furthermore, both $\mu^{long}$ and $\mu^{lat}$ were decreased by a factor of 0.70 for heavy vehicles in dry conditions. 
The availability of each occurrence alternative was included in the specification of the likelihood. A lane change conflict event, for example, is not possible in single lane sections and therefore should not be considered as an available outcome on those sections during the modeling stage. Thus, for each observation:

\begin{itemize}
    \item  a rear-end conflict is considered possible whenever the subject vehicle is in a car-following state;
	\item  a lane change conflict is considered possible if the road carriageway has two or more lanes and if the subject vehicle wants to perform a lane change;
    \item a run-off-road event is considered possible if the road section is a curve or if the subject vehicle is performing a lane change.
\end{itemize}

Ideally, the multiple replications should be used directly in the estimation phase within a Monte Carlo process, similar to panel data estimation. With this approach, several observations (replications) for the same event are available and directly included in the safety score function with an additional event specific component. The main burden in such an approach is the computer memory and processing resources needed during the estimation phase. In the current study, the estimation process was carried out considering the replications as independent. 

\subsection{Results}
Consistent estimates of the model parameters are found by maximizing the WESML function:

\begin{equation}
\mathcal{L}=\sum_{s,p}\sum_{k} y_{k,s,p} w_{k} \ln\left[P_{s,p}(k)\right]
\label{eq:loglike}
\end{equation}

where $k$ are all possible outcomes considered for the proposed model (see Figure \ref{fig:modelstruct}), $P_{s,p}(k)$ is the probability of outcome $k$ for spatial interval $s$ and time period $p$ (given by eq. \ref{eq:DiscreteProbability}), $w_k$ is the outcome $k$-specific sampling ratio, $y_{k,s,p}$ is 1 if $k$ is the observed outcome for the observation pair $s,p$ and 0 otherwise. In this study, \texttt{PythonBIOGEME} was used \cite{BIOGEME2003} for the estimation and the results are presented in Table \ref{tab:estimationresults}.
It is good practice to scale the data so that the absolute values of the parameters are between zero and 1; thus, all relative gap variation variables were divided by 10 and the lateral acceleration difference specified in 0.1$m/s^2$ units.

\begin{table}[!htb]
 \caption{Estimation results}
  \centering
  \begin{tabular}{llcccc}
    \toprule
    Event	& Parameter	& Value	& st. dev.	& t-stat	& p-val\\
    \midrule
    \multirow{4}*{Rear-end conflict RE} & $^*$RE constant $\beta_0^{RE}$	& -13.09	& 0.608	& -5.08	& <0.01\\
    & Positive relative needed dec. $\beta_1^{RE}$	& 2.917	& 0.917	& 3.18	& 0.01\\
    & Negative relative needed dec. $\beta_2^{RE}$ 	& -1.92	& 0.784	& -2.45	& 0.03\\
	& Maximum available dec. ratio $\beta_3^{RE}$ 	& 2.03	& 1.034	& 1.96	& 0.07\\
	\midrule
    \multirow{5}*{Lane-change conflict}	& $^*$LC constant $\beta_0^{LC}$	& -7.08	& 0.457	& 6.32	& <0.01\\
	& $^\dagger$Positive relative lag gap variation $\beta_1^{LC}$	& -0.011	& 0.012	& -0.92	& 0.38\\
	& Negative relative lag gap variation $\beta_2^{LC}$	& -0.568	& 0.338	& -1.68	& 0.12\\
	& $^\dagger$Positive relative lead gap variation $\beta_3^{LC}$	& -0.311†	& 0.255	& -1.22	& 0.25\\
	& Negative relative lead gap variation $\beta_4^{LC}$	& -0.628	& 0.315	& -1.99	& 0.07\\
	\midrule
	\multirow{4}*{Run-off-road event}   & $^*$ROR constant $\beta_0^{ROR}$	& -12.45	& 0.367	& -6.68	& <0.01\\
	& Positive lateral acc. difference $\beta_1^{LC}$	& 0.023	& 0.013	& 1.77	& 0.10\\
	& Negative lateral acc. difference $\beta_2^{LC}$	& 1.775	& 0.965	& 1.84	& 0.09\\
	\midrule
	\multicolumn{2}{l}{Scale parameter for the accident nest  $\mu$}    & 1.622	& 0.567	& 2.86	& 0.01\\
    \bottomrule
    \multicolumn{1}{r}{$\#$ of parameters}	&  \multicolumn{5}{l}{13 ($^*$ are the parameters affected by weights)}\\
    & \multicolumn{5}{l}{~~~~~($^\dagger$ are the parameters excluded from the final model)}\\
    \multicolumn{1}{r}{Sample size:}   & \multicolumn{5}{l}{10733084 (3 replications)}\\
	\multicolumn{1}{r}{Initial log-likelihood:}	& \multicolumn{5}{l}{-9636.49}\\
	\multicolumn{1}{r}{Final log-likelihood:}	& \multicolumn{5}{l}{-2047.53}\\
	\multicolumn{1}{r}{$\rho^2$:} & \multicolumn{5}{l}{0.787}\\
	\multicolumn{1}{r}{$\bar{\rho}^2$:}	& \multicolumn{5}{l}{0.786}
  \end{tabular}
  \label{tab:estimationresults}
\end{table}

When the $RA^{need}$ component is positive the risk of a rear-end conflict increases, as the difference between the vehicle relative deceleration rate and its \textit{DRAC} increases.  $\beta_1^{RE}$ has a higher absolute magnitude than $\beta_2^{RE}$, which penalizes safety decay in the unsafe domain ($RA^{need}>0$) rather than in the safe domain ($RA^{need}<0$). Regarding the negative component, i.e. when the follower has already adjusted his acceleration, lower $RA^{need}_{-}$ will result in an increased RE probability due to lower \textit{TTC}. The positive sign of $\beta_3^{RE}$ and its statistical significance makes the consideration of different exogenous safety conditions non-negligible. Both the vehicle category (car vs. truck/bus) and the pavement (wet vs. dry) conditions were considered.

The parameter of the negative component of the lead gap variation during LC events ($\beta_4^{LC}$) is also significant: the large absolute values of the independent variable ($RG^{lead}_{-}$) represent significantly shrinking gaps and, therefore, will increase the probability of LC accident events. The higher magnitude of the relative lead gap variation parameter is due to the much smaller simulated lead gaps during lane-change not only when compared to lag gaps but also when comparing accident events with no-accidents. The statistical significance and the estimated sign and magnitude of the negative relative lag gap variation parameter, $\beta_2^{LC}$, is consistent with the model assumptions. The parameters of the positive relative gaps variation, $\beta_1^{LC}$ and $\beta_3^{LC}$, were found non-significant and were removed from the final model used in the rest of the paper.

The safety score of an ROR event is assumed to be linked to $\delta a^{lat}$, i.e. the difference between the current lateral acceleration of a vehicle n and the site-specific critical lateral acceleration. When $\delta a^{lat}$ is positive, the lateral acceleration computed by the simulator is higher than the critical lateral acceleration and the vehicle is under unsafe conditions. Under these conditions, when $\beta_1^{ROR}>0$ there is a higher probability of ROR events. Similarly, when $\Delta a^{lat}$ is negative, larger absolute values are related to safer conditions, as the simulated lateral acceleration is much smaller than the critical one ($\beta_2^{ROR}$). One would expect a higher absolute magnitude for $\beta_1^{ROR}$, but these results may be justified with the small number of observations with $\Delta a^{lat}>0$. Furthermore, it is expected that an enhanced simulation of lateral accelerations and its integration in the safety score function in form of combined effects with the longitudinal counterparts (friction circle) would increase the model’s performance.

The estimated scale parameter of the accidents nest $\mu$ was also significant, revealing a non-negligible effect of shared unobserved attributes of the different types of accident under analysis.

\section{Validation}\label{sec:validation}

\subsection{Sensitivity to artificial data stochasticity}
In order to analyse the model sensitivity regarding the variability of the artificial trajectories, a new set of artificial data was generated in \texttt{MITSIMLab} for the same sample of events presented earlier. The proposed model was re-estimated without the less statistically significant variables. Two additional replications of the calibrated model of the A44 were carried out generating two new artificial data sets. The event-specific calibrated parameters were used in the additional replications, relying on the stochasticity parameters of the simulator for the generation of different (new) trajectories.

In Table \ref{tab:probtest} the averaged ratios of the probabilities between a specific type of accident and the no-accident events are presented for both the estimation and sensitivity data sets. The range of both input variables and estimated probabilities for the testing data set are similar to the estimation ones. The trade-offs (correlations) captured by the model are also visible, especially between the rear-end and lane-change conflicts.

\begin{table}[!htb]
 \caption{Testing probability ratios regarding $P(NA)$}
  \centering
  \begin{tabular}{llccc}
    \toprule
    & & $P(RE)$	& $P(LC)$   & $P(ROR)$\\
    \midrule
    \multirow{3}{*}{Estimation data set}    & RE	& 3.783	& 3.880	& 0.359\\
	& LC	& 2.284	& 3.581	& 0.468\\
	& ROR	& 1.755	& 0.499	& 1.241\\
	\midrule
    \multirow{3}{*}{Sensitivity data set}    & RE		& 4.352	& 5.824	& 0.344\\
	& LC	& 2.363	& 3.027	& 0.391\\
	& ROR	& 1.306	& 0.277	& 1.299\\
    \bottomrule
  \end{tabular}
  \label{tab:probtest}
\end{table}

In Table \ref{tab:prediction} the accuracy rates of the accident types considered are presented using the sensitivity data set. In a previous model using real loop sensor data, \citet{Oh2001} estimated the prediction accuracy for accidents and non-accidents as 55.8\% and 72.1\%, and a false alarm rate of 27.9\%. \citet{Xu2013a} estimated the same rates as 61.0\%, 80.0\% and 20.0\%, respectively. These magnitudes are obviously not comparable, as combined aggregated real traffic data and accident occurrences were used in model estimation, but they still represent a good indication of the state-of-the-art performances of real-time detailed accident prediction models. The rates obtained with the proposed model with artificial data still remain below the values found in the literature for aggregated accident probability models using real data (38.8\% accident prediction accuracy and 7.9\% false alarms). The small sample used for estimation may have affected this number. Yet, the false alarm rate is lower than values reported in other studies, which represents an important attribute regarding its use in real time traffic management applications. Also, the flexible nature of the model specification allows for an easy enhancement in future applications.

\begin{table}[!htb]
 \caption{Summary of predictive performance for the sensitivity data set}
  \centering
  \begin{tabular}{lcccc}
    \toprule
    & RE    & LC	& ROR	& Total Accidents\\
    \midrule
    Accuracy	& 38.8\%	& 41.1\%	& 11.6\%	& 38.8\%\\
False alarms	& 1.9\%	& 3.6\%	& 2.4\%	& 7.9\%\\
    \bottomrule
  \end{tabular}
  \label{tab:prediction}
\end{table}

\subsection{Cross validation}

As no other accident data set was available for this study, a two-fold cross validation was carried out using the original estimation data set. Half of the accident events and non-accident events were randomly assigned to the training and testing data sets of equal size (72 accidents and 3200 non-accidents, see Table \ref{tab:dataset})\footnote{The difference between the Observed and Estimation data sets is due to faulty accident records that were discarded for estimation}. The model was first estimated with the training data set and then applied to the testing set.

\begin{table}[!htb]
 \caption{Data sets}
  \centering
  \begin{tabular}{lcccc}
    \toprule
    & RE    & LC	& ROR	& NA\\
    \midrule
    Observation period (2007-2009)	& 76	& 41	& 56	& 182,600,467\\
    Estimation set	& 61	& 37	& 46	& 6,400\\
    Training set	& 36	& 15	& 21	& 3,200\\
    Testing set	    & 25	& 22	& 25	& 3,200\\
    \bottomrule
  \end{tabular}
  \label{tab:dataset}
\end{table}

The very small samples will obviously lower the performance of the validation test, as they are sensitive to the random sampling process. In Table \ref{tab:predtest} the accuracy rates of the accident types considered for the training set are presented.

\begin{table}[!htb]
 \caption{Summary of predictive performance for the testing data set}
  \centering
  \begin{tabular}{lcccc}
    \toprule
    & RE    & LC	& ROR	& Total Accidents\\
    \midrule
    Accuracy	    & 16.1\%	& 17.6\%	& 9.5\%	& 34.8\%\\
    False alarms	& 3.1\%	& 1.4\%	& 1.6\%	& 6.1\%\\
    \bottomrule
  \end{tabular}
  \label{tab:predtest}
\end{table}

\begin{figure}[!htb]
    \centering
    \includegraphics[width=.3\textwidth,frame]{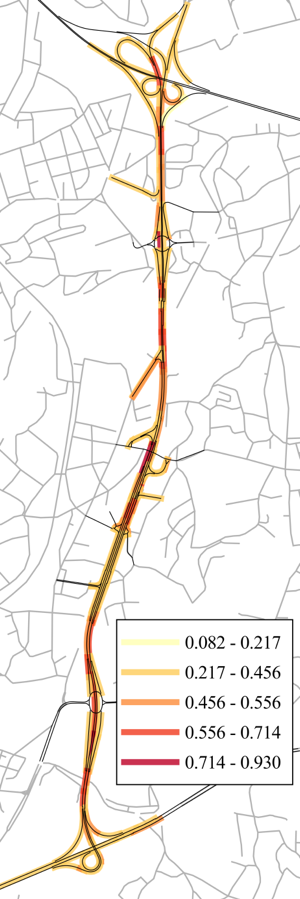}\hfill
    \includegraphics[width=.3\textwidth,frame]{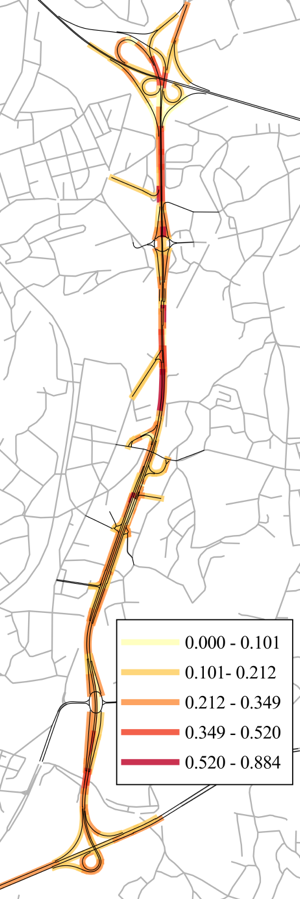}\hfill
    \includegraphics[width=.3\textwidth,frame]{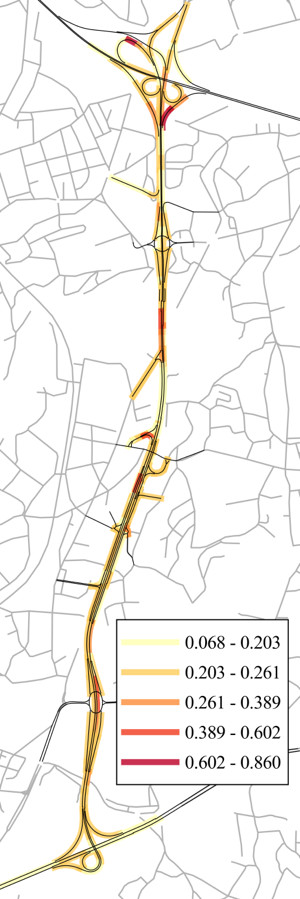}
    \caption{A44 Average probabilities (in $10^{-5}$) for RE (left), LC (centre) and ROR (right) accidents aggregated by 50m and 5 min}
    \label{fig:simul}               
\end{figure}

From the above results, the need for further validation and extension of the estimation framework to larger accident data sets is clear. As expected, the accuracy in identifying the type of accident is different from the sensitivity test results presented above, where a larger estimation data set was used. Yet, the model was able to predict a third of the real unsafe situations, showing stability when compared with the previous tests and the use of the calibrated microscopic traffic simulator helped in detailed traffic safety analysis (see Figure \ref{fig:simul}):

\begin{itemize}
    \item The probability of RE and LC events is larger in sections close to the simulated weaving areas. This is due to the dense traffic and to short interchange spacing in A44. High acceleration and deceleration along with short lateral gaps are associated with such probabilities.
	\item The upper half sections of the A44 close to entry and exit ramps have a higher probability of SC, mainly due to the absence of dedicated slip roads and sub-standard acceleration and deceleration lane lengths. \item These factors affect drivers gap acceptance, forcing lower than normal critical gaps in lane-changing manoeuvres, for vehicles to keep their routes, here captured by the calibrated simulator.
	\item Similarly, link sections following short entry ramps have higher RE event probabilities. 
	\item A few curved elements clearly show an increased probability of ROR events, revealing inadequate (simulated) speed choices in those locations.
\end{itemize}

Further research needs to be carried out to test the efficiency of the proposed method for safety assessment when comparing simulated scenarios. That is the case for adding extra lanes to diverging and weaving sections; locating variable speed limit signs and automatic enforcement systems; locating additional directional signing; and setting speed limit strategies attending to both atmospheric and traffic prevailing conditions. 

\section{Conclusions}

A generic framework for modelling cause effect mechanisms between detailed traffic variables and accident occurrence probability in traffic microscopic simulation tools was proposed and tested in a real road environment.
The safety model was detailed and estimated for an urban motorway case study, allowing for the identification and interpretation of the several vehicle interactions at stake. Rear-end accident probability is linked to the needed deceleration ratios. The maximum available deceleration ratio is also non-negligible, corroborating the importance of the consideration of different exogenous safety conditions. The lane-change conflict probability is mainly connected with shrinking lead gaps. Run-off-road events rely on the lack of available lateral accelerations. The nested structured allowed us to capture existing trade-offs between these three types of accidents. The fact that these considerations were extracted from simulated analysis shows the real potential of advanced traffic microscopic simulation regarding detailed safety assessments, as long as detailed calibration is successfully carried out. The interaction between vehicle gaps and relative motions is a key factor for accident occurrence in previous safety related studies. Yet, no probabilistic formulation accommodating such detailed interaction and integrated in traffic simulation models had previously been reported in the literature.

The availability of large detailed trajectory data sets from naturalistic studies will be a key source for potential improvements in the development of probabilistic safety models as conflict probabilities may be directly computed from trajectories collected for accidents and evasive manoeuvres. 

The most recent driving behaviour model formulations already allow for several improvements such as decoupling the simulation step from reaction times, accounting for anticipation, etc. However, even if the number of sub-models and their parameters has grown significantly, results at the disaggregated level are not always well replicated (e.g.: detailed vehicle interactions). The integration of conceptual perception and error modelling frameworks and more detailed motion descriptions in large traffic microscopic simulation tools may mitigate some of these performance constraints.

From the estimation results, it is clear that some of the variables are not significant. The small accident sample and the artificial nature of the data impact the performance of the estimated model for urban motorways. Yet, the flexible structure of the proposed model can accommodate several enhancements regarding the specific formulation of the safety score functions that may lead to increased model performance. The inclusion of further components (e.g.: driver related variables and more detailed vehicle motion and road characteristics) that might be available in new simulation models and the formulation of non-linear safety score functions are clearly the most needed improvements, especially regarding run-off-road events. The specification of additional accident types and the definition of more powerful modelling structures, such as the mixed logit, or estimation methods, such as a panel data estimation based on multiple replications, should also be tested. Moreover, the validation using other sets of data and traffic scenarios, additional statistical test when selecting optimized safety score functions and a benchmark against alternative non-probabilistic safety assessment tools would be valuable. A possible future direction is the comparison between traffic conflict analysis of the Surrogate Safety Assessment Model \cite{FHWA2008} and the proposed framework in the safety analysis of simulated scenarios.

Finally, it is important to point out that the integration of such safety model into simulation tools and its use for traffic management policy implementation in very different scenarios requires further estimation and validation tasks, with other case-studies, as scant information was available for this study. The flexible structure and methodology demonstrated in this document allows for a valid and consistent assessment of accident occurrence for specific driving behaviour models. It is worth remembering that the modelling and estimation structures were formulated in terms of expected behavioural considerations but constrained by the driving behaviour simulation model limitations. When a safety assessment model (probabilistic or not) is integrated into a simulation tool, the safety formulation should also consider the modelling assumptions and limitations of the traffic simulator.

\section{Acknowledgements}
This research was supported by the Portuguese National Funding Agency for Science, Technology and Innovation (\textit{Funda\c{c}\~{a}o para a Ci\^{e}ncia e Tecnologia}) and the Portuguese National Laboratory for Civil Engineering, with the PhD scholarship award number \texttt{SFRH/BD/35243/2007} under the Massachusetts Institute of Technology - Portugal Program. The research presented also benefited from the computational resources of the Portuguese National Grid Initiative (\url{https://wiki.ncg.ingrid.pt}).


\bibliographystyle{unsrtnat}

\end{document}